# On-Surface Synthesis of Silole and Disilacyclooctaene Derivatives

Kewei Sun[1]*, Lauri Kurki[3], Orlando J. Silveira[3], Tomohiko Nishiuchi[4], Takashi Kubo[4], Ondřej Krejčí[3], Adam S. Foster[3,5]*, Shigeki Kawai[2,6]*

[1]*International Center for Young Scientists, National Institute for Materials Science, 1-2-1 Sengen, Tsukuba, Ibaraki 305-0047, Japan.*

[2]*Center for Basic Research on Materials, National Institute for Materials Science, 1-2-1 Sengen, Tsukuba, Ibaraki 305-0047, Japan.*

[3]*Department of Applied Physics, Aalto University, Espoo, Finland.*

[4]*Department of Chemistry, Graduate School of Science, Osaka University, Toyonaka 560-0043, Japan.*

[5] *Nano Life Science Institute (WPI-NanoLSI), Kanazawa University, Kakuma-machi, Japan.*

[6]*Graduate School of Pure and Applied Sciences, University of Tsukuba, Tsukuba 305-8571, Japan.*

**Abstract**

Sila-cyclic rings are a class of organosilicon cyclic compounds and have abundant application in organic chemistry and materials science. However, it is still challenging to synthesize compounds with sila-cyclic rings in solution chemistry due to their low solubility and high reactivity. Recently, on-surface synthesis was introduced into organosilicon chemistry as 1,4-disilabenzene bridged nanostructures were obtained via coupling between bromo-substituted molecules and silicon atoms on Au(111). Here, we extend this strategy for syntheses of silole derivatives and graphene nanoribbons with eight-membered sila-cyclic rings from 2,2',6,6'-tetrabromobiphenyl and 1,4,5,8-tetrabromonaphthalene on Au(111), respectively. Their structures and electronic properties were investigated by a combination of scanning tunneling microscopy/spectroscopy and density functional theory calculations. This work demonstrates a generality of this synthesis strategy to fabricate various silicon incorporated nanostructures.

**Introduction**

Organosilicon chemistry deals with syntheses and characterization of compounds containing carbon-silicon (C-Si) bonds. Ever since the first synthesis of tetraethylsilane by Friedel and Crafts in 1863,[1,2] various synthetic strategies have been developed for the molecules with C-Si bonds, such as silyl ethers,[3] silyl chlorides,[4] silenes[5] and siloles.[6] Among them, the silicon incorporated cyclic ring, namely sila-cyclic ring, has attracted the attention of researchers and engineers[6-9] because the sila-cyclic ring can be used for various applications such as catalysis,[10,11] precursors of molecular transformation,[12,13] electron transport[14,15] and light-emitting materials.[16,17] Although several sila-cyclic rings have been synthesized,[6-9] it is still challenging to incorporate them in polycyclic aromatic hydrocarbons due to their low solubility and high reactivity.

On-surface chemistry deals with the important study of chemical reactions and characterization of molecules on solid substrates. In the reaction, small organic molecules adsorbed on metal surfaces are usually activated by annealing, irradiated light, and injected tunneling electrons and subsequently conjugated with each other. The advantage of on-surface synthesis relates to the high-controllability of the structures, which can be defined by the employed precursor molecules.[18] So far, various surface reactions have been demonstrated, such as Ullmann-type coupling,[19-21] Glaser-type coupling,[22-24] Bergman-type reaction,[25,26] Sonogashira-type coupling,[27-29] and dehydrogenated coupling.[30,31] Very recently, C-Si bond coupling was obtained by reacting bromo-substituted molecules to Si atoms directly on Au(111), leading to the formation of 1,4-disilabenzene bridged covalent organic frameworks (COFs) and graphene nanoribbons (GNRs).[32] This strategy is expected to have a high generality for the synthesis of multiple sila-cyclic rings with appropriate precursors.

Here, we demonstrate the on-surface synthesis of five and eight-membered sila-cyclic rings on Au(111). 2,2',6,6'-tetrabromobiphenyl **1** having two bromines (Br) at the *bay* position reacted with Si atoms to form two siloles ($C_4Si$) incorporated into a molecule at 420 K. For 1,4,5,8-tetrabromonaphthalene **2** having two Br at the *peri* position, corrugated GNRs embedded with eight-membered sila-cyclic rings ($C_6Si_2$) were generated by reacting with Si atoms after annealing at 470 K. Each Si atom in the cyclic rings was passivated by one Br atom, which was then removed by high-temperature annealing. The structure and the electronic properties of the products were investigated with a combination of scanning tunneling microscopy/spectroscopy (STM/STS) and density functional theory (DFT) calculations.

**Results and discussions**

In our previous study, two Br atoms at *ortho* position of peripheral phenyl groups in 2,3,6,7,10,11-hexabromotriphenylene reacted to Si atoms adsorbed on Au(111), yielding 1,4-disilabenzene ($C_4Si_2$ ring) bridged nanostructures (Scheme 1a).[32] Following this concept, two Br atoms at *bay* (Scheme 1b) and *peri* positions (Scheme 1c) are used for syntheses of five-membered sila-cyclic ring ($C_4Si$) and eight-membered sila-cyclic ring ($C_6Si_2$) by conjugating **1** and **2** with Si atoms on Au(111), respectively (Scheme 1d).

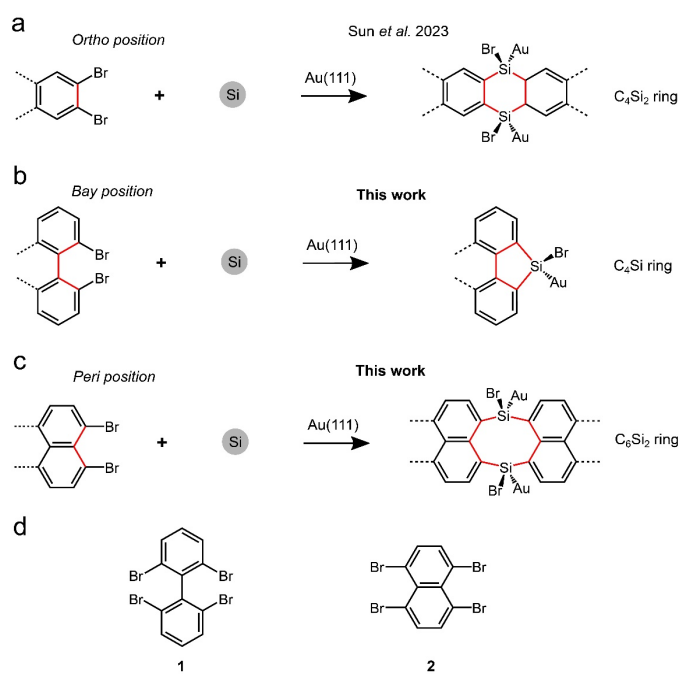

**Scheme 1**. On-surface reactions between bromo-substituted aromatic hydrocarbons and Si atoms on Au(111). Two Br atoms at (**a**) *ortho* position of phenyl group: synthesis of 1,4-disilabenzene ($C_4Si_2$) bridged nanostructures (previous work)[32]; (**b**) *bay* position of biphenyl group: synthesis of five-membered ring ($C_4Si$); (**c**) *peri* position of naphthyl group: synthesis of eight-membered ring ($C_6Si_2$) bridged structures. (**d**) **1** and **2** as precursors.

We first deposited Si atoms on a clean Au(111) surface and subsequently annealed the substrate at 420 K to form the AuSi$_x$ submonolayer.[32-34] After depositing **1** on the surface kept at room temperature, the sample was heated at 420 K for 5 min. Small clusters indicated by arrows in Figure 1a correspond to silicon bromide $SiBr_x$ (x = 1, 2, 3) molecules, which is in agreement with our previous work.[35] Thus, the partially covered $AuSi_x$ layer acts as a source of Si atoms for on-surface reaction.[32,35] We also found isolated molecules with two bright dots marked by dashed squares. Since such features are absent in the on-surface synthesis only with **1** on Au(111) (Figure S1), the bright dots should relate to Si atoms. We found three different types of products (Figure 1a). Among them, Type 1 has the highest chemoselectivity of 86% (Figure S2). The close-up view shows two bright dots at both right- and left-hand sides of the molecule (Figure 1b). To investigate the inner structure, the STM tip apex was terminated by a CO molecule.[36,37] The high-resolution constant height d$I$/d$V$ map (Figure 1c, Figure S3) and the corresponding Laplace filtered image (Figure 1d) show the biphenyl backbone at the middle and two sets of five-membered rings at both sides. Thus, debrominated **1** was connected to two Si atoms, resulting in formation of silole rings (Figure 1e). We also found each Si atom terminated by one Br. A similar reaction at the *ortho* position was observed in the formation of 1,4-disilabenzene.[32] Our DFT electron density of the Type 1 deposited on a bare Au(111) surface showed that the silicon atoms adsorb on top positions and bond significantly with the substrate gold atoms, slightly buckling the molecule due to an elevation of the bromine atoms (Figure 1f). We also performed STM and d$I$/d$V$ simulations which confirmed that the sharp line appearing near the C-Si bond (Figure 1d) is not a direct observation of the chemical bond, but it is rather caused by mechanical bending of the functionalized CO tip (Figure 1h). The C-Si bond length was 1.90 Å which is a typical length for a C-Si single bond.[38] The structures of Type 2 and Type 3 were also investigated with bond-resolved STM. Type 2 is longer than Type 1 in length (Figure 1i). The corresponding bond-resolved image shows that the structure is composed of two naphthyl groups and silole rings (Figure 1j). We assume that the molecular units are connected to each other via a Si-Si bond (Figure 1k), which is further illustrated by DFT simulation (inset of Figure 1j). Type 3 corresponds to the single unit, yet one Br seems to be replaced by one hydrogen atom (Figure S4).

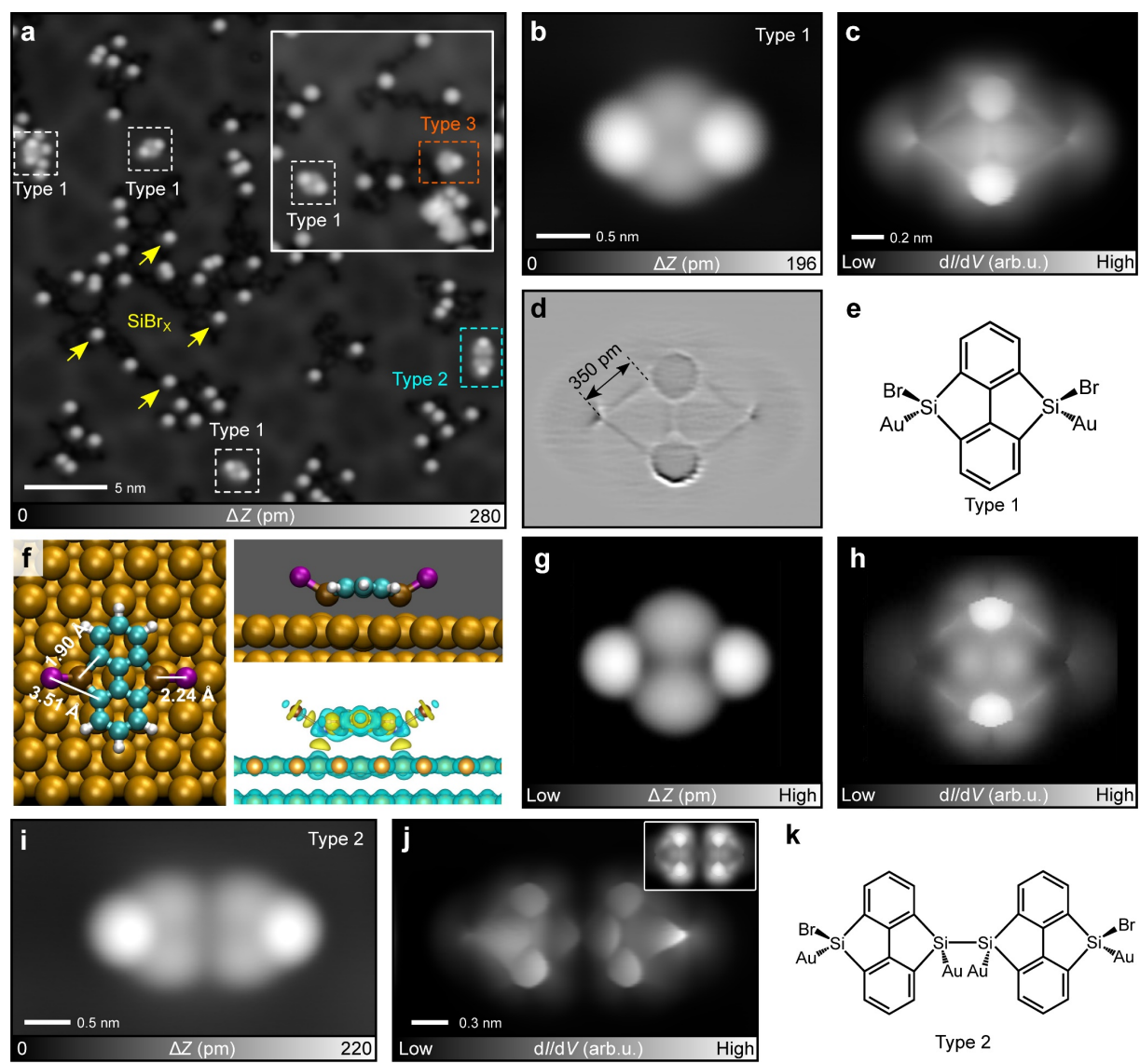

**Figure 1.** Synthesis of five-membered sila-cyclic rings from **1** on Au(111). (**a**) STM topography of Type 1 and Type 2 molecules as well as SiBr$_x$ clusters obtained by annealing at 420 K for 5 min. The inset shows Type 3 found in another area. (**b**) Close-up view of Type 1, (**c**) the bond-resolved d$I$/d$V$ map, and (**d**) the corresponding Laplace filtered image. (**e**) Chemical structure of Type 1. (**f**) Top (left) and side view (right-top) of DFT optimized structure of Type 1 on Au(111) and the electron density difference (right-bottom), (**g**) simulated STM image, and (**h**) simulated bond-resolved image. (**i**) Close-up view of Type 2 and (**j**) the corresponding constant height d$I$/d$V$ map. The inset shows the simulated bond-resolved image. (**k**) Chemical structure of Type 2. Measurement parameters: Sample bias voltage $V$ = 200 mV and tunneling current $I$ = 5 pA in (a) and (b), $V$ = 100 mV and $I$ = 20 pA in the inset of (a), $V$ = 200 mV and $I$ = 5 pA in (i), and $V$ = 1 mV and $V_{ac}$ = 10 mV in (c),(j).

Next, the electronic properties of Type 1 were measured by STS (Figure 2a). The differential conductance d$I$/d$V$ spectra were recorded at four different sites of Type 1 and one site on the bare Au(111) surface for the reference as indicated by dots in the inset of Figure 2a. Besides the Au surface state appearing around -0.5 V, two characteristic peaks at -0.6 V and 2.5 V with respect to the Fermi level were identified as the highest occupied molecular orbital (HOMO) and the lowest unoccupied molecular orbital (LUMO), respectively. Thus, the HOMO-LUMO gap of Type 1 was approximately 3.1 eV. The spatial distributions of these electronic states are seen in the constant height d$I$/d$V$ maps recorded at bias voltages of -0.6 V and 2.5 V (Figure 2b, 2c). We found that the HOMO state is located at the silole rings while the LUMO state exhibits much higher intensity at the biphenyl backbone. Similar electronic states were also observed in the 1,4-disilabenzene.[32] Constant current d$I$/d$V$ maps also show the similar bright features (Figure S5). The DFT calculated projected density of states (pDOS) of Type 1 adsorbed on a bare Au(111) surface exhibits overall similar features compared to the STS measurement (Figure 2d). In the occupied region, the contribution from the bromine atoms near –1.1 eV increases although the absolute magnitude of pDOS contribution from all carbon atoms was greater. In the unoccupied region, increasing pDOS starting at around +1.8 eV coming from carbon and silicon atoms was observed, which then peaks at around +2.1 eV. The local density of states (LDOS) of all states between –1.5 and 2.5 eV were carefully checked, which allowed the identification of HOMO and LUMO-like orbitals of Type 1 on Au(111) (Figure 2e, 2f). Between these two regions, the pDOS of the molecule stayed mostly constant, apart from small contributions from the Si atoms (further confirming that there is hybridization between Si and Au atoms), which matches with the clean band gap observed in the experiment. Additionally, we conducted constant height d$I$/d$V$ simulations, in which the bromine atoms dominate at both HOMO and LUMO energies (Figure 2g, 2h). At the HOMO energy, the signal intensity above the biphenyl backbone was larger at the top and bottom edges of the biphenyl whereas the contrast was more equally distributed at the LUMO energy, and a similar observation was also seen in the experiment (Figure 2b, 2c).

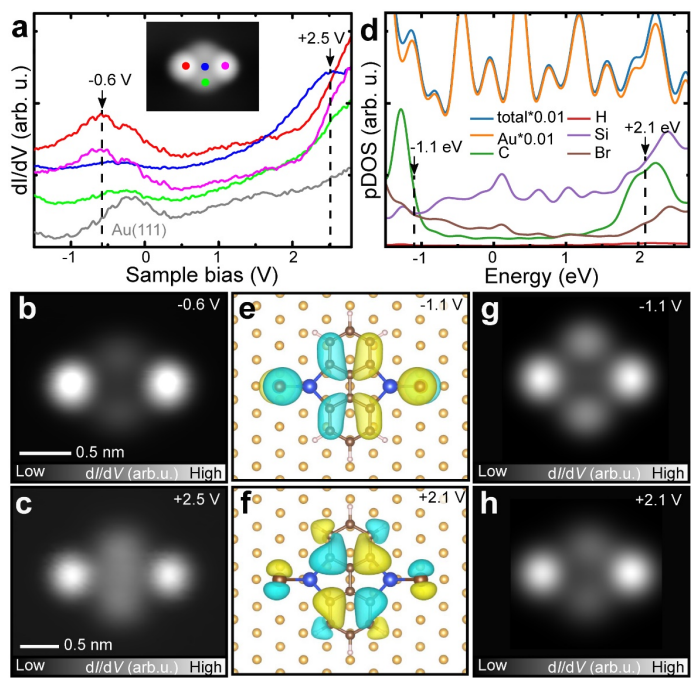

**Figure 2.** Electronic properties of Type 1 molecule. (**a**) d$I$/d$V$ curves measured at the four sites of Type 1 and on the bare Au(111) surface as indicated by dots in the inset. Constant height d$I$/d$V$ maps measured at bias voltages of (**b**) -0.6 V and (**c**) 2.5 V. (**d**) Calculated density of state of Type 1. (**e-h**) Calculated orbitals and simulated constant height images at –1.1 V and +2.1 V. Measurement parameters: $V$ = -0.6 V, $V_{ac}$ = 10 mV in (b). $V$ = 2.5 V, $V_{ac}$ = 10 mV in (c).

In contrast to **1**, **2** has two Br atoms at the *peri* position at each side. We deposited **2** on Au(111) partially covered with the AuSi$_x$ layer at room temperature, subsequently annealing at 470 K for 5 min. Similar to the reaction with **1**, the AuSi$_x$ layer disappears, and consequently the bare Au(111) surface was restored. We observed the formation of one-dimensional structures on the surface (Figure 3a). The close-up view shows that the shape of the product differs from those of organometallic chains[21,39] and N = 5 armchair GNRs synthesized with **2** alone on a clean Au(111) surface (inset of Figure 3a).[40] The bright dots in the one-dimensional structures are located at periodic sites with a gap of 1.10 ± 0.01 nm while the apparent STM heights vary (Figure 3b). The large corrugation amplitude of the bright dots prevented high-resolution imaging in constant height mode because the CO molecule on the tip apex was often detached due to excessive interactions. Thus, we attempted to resolve only the molecular backbone by setting a narrow scan area as indicated by a square in Figure 3b. Although it is not clear in the d*I*/d*V* map (Figure 3c), we could see the perylene backbone, which was formed via debrominative homo-coupling of **2** (Figure 3c, d). Since the longitudinal axis of the perylene is 0.7 nm in length, the bright dot results in a gap of 0.4 nm. Given the fact that the length of a C-Si bond is in a range of 0.18-0.20 nm,[32,38,41,42] we assigned a C-Si-C bond as the linker. The incorporated Si atom was passivated by one Br atom, appearing as the bright dot in the STM topography.[32] Since two dots were seen between the perylene units, two sets of C-Si-C bridges should exist, indicating formation of a $C_6Si_2Br_2$ ring, the so-called eight-membered sila-cyclic ring bridged GNR (Figure 3e). To verify our assignment of the structure, we conducted DFT calculations of the ribbon structure adsorbed on bare Au(111). Again, we saw bonding between the silicon and gold atoms and similar orientation of the Si-Br bond compared to Type 1 (Figure 3e), and STM simulations confirmed that the bright dots can be attributed to the bromine atoms (Figure 3f,3g). We constructed the DFT unit cell to include two perylene and Si-Br pairs to investigate the varying brightness of bromine atoms, but the optimized structure is mostly planar, and the silicon atoms adsorb on equivalent top sites resulting in no difference in brightness. The elongated C-Si bonds (0.2 nm) were attributed to the slightly increased lattice constant of the GNR to achieve commensurability with the substrate. To investigate the reason for the apparent alternation we forced one pair of bromine atoms to relax 0.02 nm lower than the adjacent pair and the simulated STM image of this structure showed alternating brightness (Figure 4j). Therefore, we assign the contrast difference to different silicon adsorption sites affecting the bromine height.

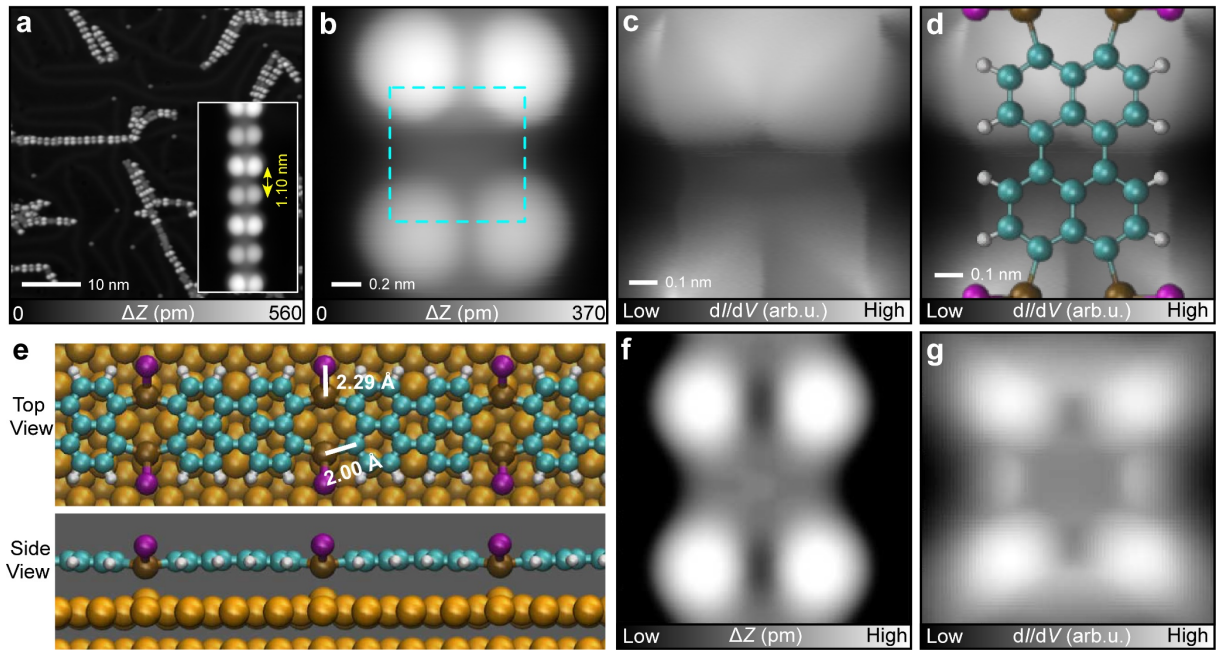

**Figure 3.** Synthesis of eight-membered sila-cyclic rings bridged GNRs. (**a**) Large-scale STM topography of the surface after annealing at 470 K for 5 min. Inset shows the STM topography of individual nanoribbon. (**b**) Close-up view of one unit in a nanoribbon. (**c**) High-resolution constant height d$I$/d$V$ map of the area indicated by dashed square in (b). (**d**) Chemical model superimposed on (c). Optimized structure of the nanoribbon (**e**). Simulated constant current STM image, simulated bond-resolved d$I$/d$V$ map (**f, g**). Measurement parameters: $V$ = 200 mV and $I$ = 2 pA in (a). $V$ = 200 mV and $I$ = 5 pA in (b). $V$ = 1 mV, $V_{ac}$ = 10 mV in (c).

To investigate the electronic properties of eight-membered sila-cyclic ring bridged GNR, we performed STS measurements focusing on a particular case where the ribbon exhibits periodically undulating brightness. Compared with the result on the bare Au(111) surface (grey curve), we found three distinct peaks at -0.6 V, 1.9 V and 2.5 V, which were measured above three different sites in the GNR (Figure 4a). The two bright dots with different contrasts exhibit distinct empty states at 1.9 V and 2.5 V, which may be related to their adsorption heights because the substrate polarization effect enhanced by a strong adsorption shifts the empty state towards the Fermi level.[43,44] The occupied state was visible at -0.6 V only above the dots with the higher adsorption height in GNR. Nevertheless, the band gap of this GNR is 2.5 eV. The STM topographies and the simultaneously recorded d$I$/d$V$ maps obtained at bias voltages of -0.6 V, 1.9 V and 2.5 V show that the three electronic states mainly distribute on the sites of eight-membered rings (Figure 4b-g). We calculated the electronic structure of the GNR using the structure obtained by constraining vertical relaxation of the bromine atoms, and the calculated pDOS of the GNR rapidly increased in the unoccupied region, with energies larger than +1.9 eV peaking at approximately +2.5 eV (Figure 4h). In the occupied region there was a brief increase in pDOS between –0.5 eV and –1.0 eV. Simulated constant current STM images were mostly dominated by the lobes located at bromine atom sites, appearing larger with positive energies (Figure 4i, 4k, 4m). The orbital structures at –0.6 eV, +1.9 eV and +2.5 eV corresponded with the STM images, showing that the lobes were the result of bromine atoms orienting upwards from the substrate (Figure 4j, 4l, 4n).

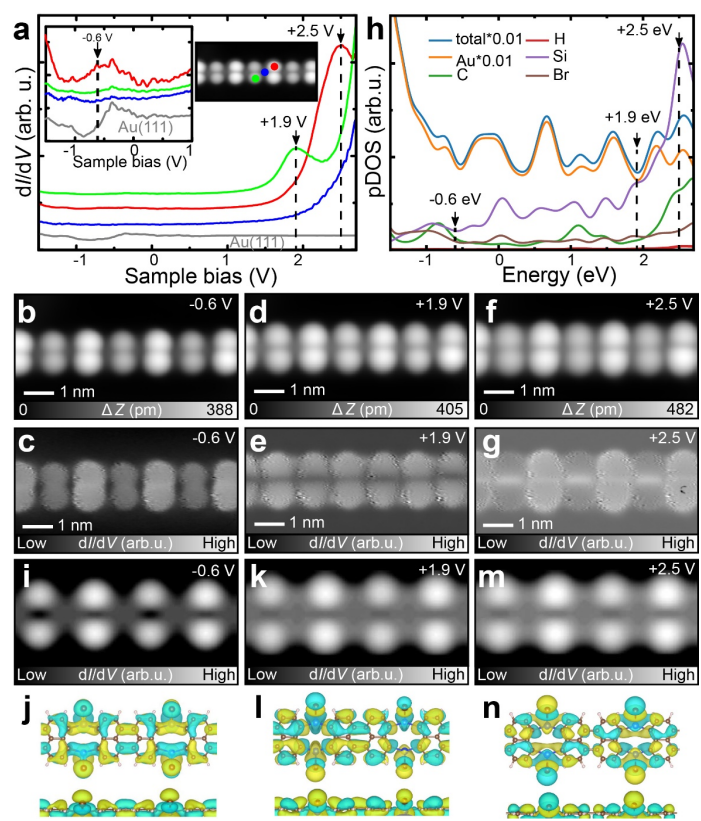

**Figure 4.** Electronic properties of $C_6Si_2$ sila-cyclic ring bridged GNR. (**a**) d$I$/d$V$ curves recorded on the GNR (inset of (a) and the bare Au(111) substrate. (**b-g**) Constant current STM topographies and the corresponding d$I$/d$V$ maps measured at sample bias voltages of -0.6 V, 1.9V and 2.5 V, respectively. (**h**) Calculated density of states of the eight-membered sila-cyclic ring bridged GNR and (**i-n**) the corresponding constant current STM simulations and orbitals of occupied and unoccupied states in top and side views. Measurement parameters: $V$ = -0.6 V, $I$ = 150 pA in (b). $V$ = -0.6 V, $I$ = 150 pA, $V_{ac}$ = 10 mV in (c). $V$ = 1.9 V, $I$ = 220 pA in (d). $V$ =

1.9 V, $I$ = 220 pA, $V_{ac}$ = 10 mV in (e). $V$ = 2.5 V, $I$ = 220 pA in (f). $V$ = 2.5 V, $I$ = 220 pA, $V_{ac}$ = 10 mV in (g).

In addition to the $C_6Si_2$ rings, we also found other sila-cyclic rings. Figure 5a shows an example, which has a planar segment as indicated by an arrow (also see Figure S6). The close-up view indicates that the segment has two faint dots (inset of Figure 5a). The corresponding bond-resolved constant height d$I$/d$V$ map (Figure 5b) and Laplace filtered image (Figure S7) revealed the detailed structure as two silole rings incorporated into a GNR segment (Figure 5c). The $C_5Si$ ring was most probably formed through sequential C-Si activation, C-C coupling, dehydrogenation and a subsequent C-Si coupling process of the $C_6Si_2$ rings (Figure S8). Figure 5d shows another type of the planar segment, in which two dots attached to the edges of GNR as indicated by an arrow. The constant height d$I$/d$V$ map (Figure 5e) and Laplace filtered image (Figure S9) indicate the formation of an N = 5 armchair GNR segment incorporated by two five-membered sila-cyclic rings ($C_4Si$) (Figure 5f). This segment was most probably generated through sequential transformation reactions of the $C_6Si_2$ rings (Figure S10). Finally, flat GNRs with dark lines, running across the longitudinal axis, were found (Figure 5g). A close-up view shows one GNR segment separated by dark line as indicated by an arrow (inset of Figure 5g). The bond-resolved STM image (Figure 5h) and the corresponding Laplace filtered image (Figure S11) revealed that each segment corresponds to the perylene unit. However, we found that the structure around the dark line cannot be resolved by constant height d$I$/d$V$ mapping. Given the fact that the individual Si atom is strongly adsorbed to the Au(111) surface,[32] the dark lines should relate to the presence of the Si atom (Figure 5i). Since there was no bright dot in the STM topography and no characteristic contrast in the d$I$/d$V$ map, it is likely that Br atoms have already desorbed from the Si atom in the product. Thus, we deduce that the perylene units were linked by an $C_6Si_2$ ring. In fact, high-temperature annealing at 620 K induced removal of the bright spots from the GNRs (Figure 5j, Figure S12).

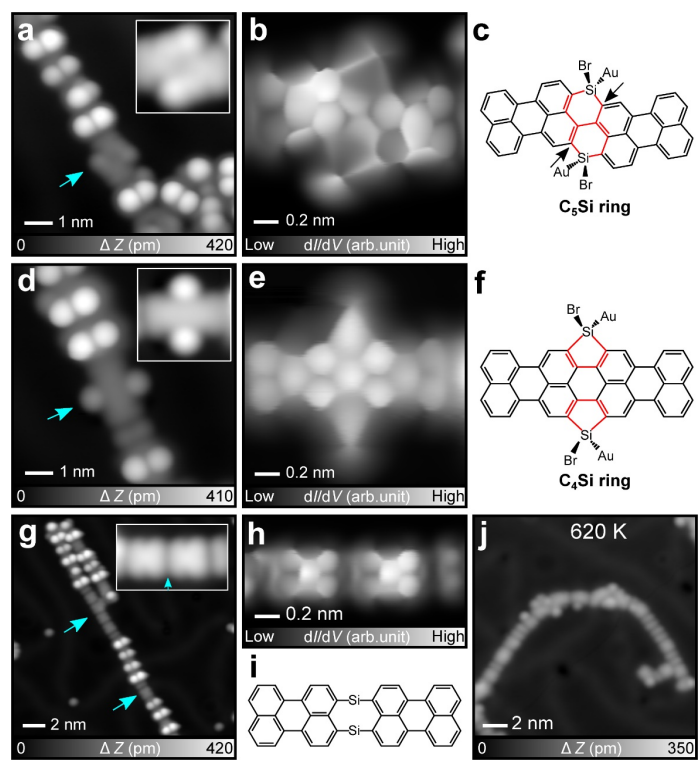

**Figure 5.** Various sila-cyclic rings incorporated in GNRs. (**a**) STM topography of one GNR. Inset shows the segment indicated by an arrow in (a). (**b**) High-resolution constant height d$I$/d$V$ map of the segment, and (**c**) the corresponding chemical structure. (**d**) STM topography of another GNR. Inset shows the segment indicated by arrow in (d). (**e**) High-resolution constant height d$I$/d$V$ map of the segment, and (**f**) the corresponding chemical structure. (**g**) STM topography of two GNRs. Inset shows the segment with dark line. (**h**) High-resolution constant height d$I$/d$V$ map of the segment. (**i**) Chemical model of debromination in $C_6Si_2Br_2$ moieties between two perylene. (**j**) STM topography of the sample after annealing at 620 K for 5 min. Measurement parameters: $V$ = 200 mV and $I$ = 2 pA in (a), $V$ = 200 mV and $I$ = 5 pA in (d). $V$ = 200 mV and $I$ = 4 pA in (g), and $V$ = 1 mV and $V_{ac}$ = 10 mV in (b),(e),(h).

**Conclusion**

We fabricated two five-membered sila-cyclic rings ($C_4Si$) incorporated molecule Type 1 and eight-membered rings ($C_6Si_2$) bridged GNRs on Au(111) by reacting Si atoms with molecules **1** and **2**, respectively. Their chemical structures were revealed by bond-resolved STM and DFT calculations. A HOMO-LUMO gap of 3.1 eV was determined for Type 1 molecules and a band gap of 2.5 eV was assigned to eight-membered rings bridged GNR by STS measurements. This research demonstrates the robustness of a synthetic strategy to obtain low-dimensional Si incorporated nanostructures by direct on-surface coupling. We foresee that this method has a potential to be extended to the synthesis nanostructure incorporating heavier elements, such as germanium, allowing for even more flexibility over the resulting electronic structure.

**Methods**

STM experiments: All experiments were performed in a low temperature scanning tunneling microscopy (STM) system (home-made) at 4.3 K under ultrahigh vacuum condition ($< 5 \times 10^{-10}$ mbar). A clean single crystal Au(111) substrate was prepared through cyclic sputtering ($Ar^+$, 10 min) and annealing (730 K, 15 min). The temperature of the sample was measured by a thermocouple and a pyrometer. 2,2',6,6'-tetrabromobiphenyl and 1,4,5,8-tetrabromonaphthalene molecules were deposited from Knudsen cells (Kentax GmbH). To form $AuSi_x$ sub monolayer, silicon atoms were deposited on a clean Au(111) surface with an electron beam evaporator (SPECS GmbH). A STM tip was made from the chemically etched tungsten. For constant height $dI/dV$ imaging, the tip apex was terminated by a CO molecule picked up from the surface. The bias voltage was set close to zero voltage. The modulation amplitude was 7 $mV_{rms}$ and the frequency was 510 Hz.

Synthesis of molecule: Compound **1** and **2** were synthesized by following reported procedures refs. 45 and 40, respectively.

Theoretical calculations: All ab-initio calculations were performed using the FHI-aims code with the PBE functional and the $vdW^{surf}$ method to include the van der Waals interactions. The systems were optimized using three substrate Au layers allowing the top two layers and the adsorbate to relax until the minimum absolute force component acting on any atom was less than 0.01 eV/Å. A 3x3x1 k-grid and the 'light' basis set were used in the geometry relaxation.

A single point calculation with a gamma k-point was performed using the optimized geometry to calculate the band structure used in the STM and d$I$/d$V$ simulations which were all acquired using the PP-STM code.[46] In the STM simulations, a broadening parameter of 0.3 eV was used and the CO tip was approximated by 13 % of the current coming through the *s*-channel and 87 % coming through the $p_{xy}$-channel.[32,47] The simulation software produces constant height (CH) STM and d$I$/d$V$ images, and the simulated constant current (CC) images were calculated as CC isosurfaces from the CH images. Atomic and electronic structure visualizations were made using XCRYSDEN[48], VESTA[49] and VMD.[50]

**Data availability**

The authors declare that the data supporting the findings of this study are available within the paper and its Supplementary Information.

**Acknowledgements**

This work was supported in part by Japan Society for the Promotion of Science (JSPS) KAKENHI Grant Number 22H00285. Kewei Sun acknowledges the supporting of ICYS project. The authors acknowledge funding from the Academy of Finland (project no. 346824). A.S.F. was supported by the World Premier International Research Center Initiative (WPI), MEXT, Japan. L.K. acknowledges funding from the Finnish Cultural Foundation. The authors acknowledge the computational resources provided by the Aalto Science-IT project and CSC, Helsinki.

**AUTHOR INFORMATION**

**Contributions**

K.S. and S.K. conceived the project. K.S. performed the experiment. L.K., O.J.S., O.K. and A.S.F. performed the DFT calculations. T.N. and T.K. synthesized the precursor molecules. K.S., L.K. and S.K. wrote the paper with input from all the other authors.

**Corresponding Authors**

*SUN.Kewei@nims.go.jp

*adam.foster@aalto.fi

*KAWAI.Shigeki@nims.go.jp

**Competing interests:** The authors declare that they have no competing interests.

**Notes**

The authors declare no competing financial interest.

**Supporting Information**

Molecules **1** on Au(111), Statistical analysis of chemoselectivity comparison for three types of products, High-resolution images of Type 1, Type 3 on Au(111), STS measurement on Type 1, More GNRs on Au(111), $C_5Si$ rings on Au(111), Proposed reaction pathway for the formation of two $C_5Si$ cyclic rings incorporated GNR segment, $C_4Si$ rings on Au(111), Proposed reaction pathway for the formation of two $C_4Si$ cyclic rings incorporated GNR segment, GNR segment on Au(111), A series of STM topographies of the sample after annealing at 620 K for 5 min (PDF).

**References**

1. Friedel, C. & Crafts, J. M. Ueber einige new organische Verbindungen des Siliciums und das Atomgewicht dieses Elementes. *Liebigs Ann. Chem.*, **127**, 28−32 (1863).
2. Muller, R. One hundred years of organosilicon chemistry. *J. Chem. Educ.*, **42**, 41−47 (1965).
3. Brook, A. G. Isomerism of some α-hydroxy silanes to silyl ethers. *J. Am. Chem. Soc.*, **80**, 1886−1889 (1958).

4. Di Giorgio, P. A., Sommer, L. H. & Whitmore, F. C. Complete chlorination of methyltrichlorosilane. *J. Am. Chem. Soc.*, **70**, 3512–3514 (1948).

5. Brook, A. G., Abdesaken, F., Gutekunst, B., Gutekunst, G. & Kallury, R. K. A solid silaethene: isolation and characterization. *J.C.S. Chem. Comm.*, 191–192 (1981).

6. Rüuhlmann, K., Hagen, V. & Schiller, K. Darstellung und reaktionen der silole. *Z. Chem.*, **7**, 353–354 (1967).

7. Tokotoh, N. New progress in the chemistry of stable metallaaromatic compounds of heavier group 14 elements. *Acc. Chem. Res.*, **37**, 86–94 (2004).

8. Okumura, S., Sun, F., Ishida, N. & Murakami, M. Palladium-catalyzed intermolecular exchange between C−C and C−Si α-bonds. *J. Am. Chem. Soc.*, **139**, 12414−12417 (2017).

9. Zhao, W., Gao, F. & Zhao, D. Intermolecular s-bond cross-exchange reaction between cyclopropenones and (benzo)silacyclobutanes: straightforward access towards sila(benzo)cycloheptenones. *Angew. Chem. Int. Ed.*, **57**, 6329–6332 (2018).

10. Kubota, K., Hamblett, C. L., Wang, X. & Leighton, J. L. Strained silacycle-catalyzed asymmetric Diels–Alder cycloadditions: the first highly enantioselective silicon Lewis acid catalyst. *Tetrahedron*, **62**, 11397–11401 (2006).

11. Carlson, P. R., Burns, A. S., Shimizu, E. A., Wang, S. & Rychnovsky, S. D. Silacycle-templated intramolecular Diels−Alder cyclizations for the diastereoselective construction of complex carbon skeletons. *Org. Lett.*, **23**, 2183−2188 (2021).

12. Birot, M., Pilot, J. P. & Dunogues, J. Comprehensive chemistry of poiycarbosilanes, polysilazanes, and polycarbosilazanes as precursors of ceramics. *Chem. Rev.*, **95**, 1443−1477 (1995).

13. Mu, Q., Chen, J., Xia, C. & Xu, L. Synthesis of silacyclobutanes and their catalytic transformations enabled by transition-metal complexes. *Coord. Chem. Rev.*, **374**, 93−113 (2018).

14. Tamao, K., Uchida, M., Izumizawa, T., Furukawa, K. & Yamaguchi, S. Silole derivatives as efficient electron transporting materials. *J. Am. Chem. Soc.*, **118**, 11974−11975 (1996).

15. Uchida, M., Izumizawa, T., Nakano, T., Yamaguchi, S., Tamao, K. & Furukawa, K. Structural optimization of 2,5-diarylsiloles as excellent electron-transporting materials for organic electroluminescent devices. *Chem. Mater.*, **13**, 2680−2683 (2001).

16. Luo, J., Xie, Z., Lam, J. W. Y., Cheng, L., Chen, H., Qiu, C., Kwok, H. S., Zhan, X., Liu, Y., Zhu, D. & Tang, B. Z. Aggregation-induced emission of 1-methyl-1,2,3,4,5-pentaphenylsilole. *Chem. Commun.*, 1740–1741 (2001).

17. Zhao, Z., He, B. & Tang, B. Z. Aggregation-induced emission of siloles. *Chem. Sci.*, **6**, 5347–5365 (2015).

18. Grill, L. & Hecht, S. Covalent on-surface polymerization. *Nat. Chem.*, **12**, 115−130 (2020).

19. Hla, S., Bartels, L., Meyer, G. & Rieder, K. Inducing all steps of a chemical reaction with the scanning tunneling microscope tip: towards single molecule engineering. *Phys. Rev. Lett.*, **85**, 2777−2780 (2000).

20. Grill, L., Dyer, M., Lafferentz, L., Persson, M., Peters, M. V. & Hecht, S. Nano-architectures by covalent assembly of molecular building blocks. *Nat. Nanotechnol.*, **2**, 687−691 (2007).

21. Wang, W., Shi, X., Wang, S., Van Hove, M. A. & Lin, N. Single-molecule resolution of an organometallic intermediate in a surface-supported Ullmann coupling reaction. *J. Am. Chem. Soc.*, **133**, 13264−13267 (2011).

22. Zhang, Y.-Q., Kepčija, N., Kleinschrodt, M., Diller, K., Fischer, S., Papageorgiou, A. C., Allegretti, F., Björk, J., Klyatskaya, S., Klappenberger, F., Ruben, M. & Barth, J. V. Homo-coupling of terminal alkynes on a noble metal surface. *Nat. Commun.*, **3**, 1286 (2012).

23. Gao, H.-Y., Wagner, H., Zhong, D., Franke, J.-H., Studer, A. & Fuchs, H. Glaser coupling at metal surfaces. *Angew. Chem. Int. Ed.*, **52**, 4024−4028 (2013).

24. Kawai, S., Krejčí, O., Foster, A. S., Pawlak, R., Xu, F., Peng, L., Orita, A. & Meyer, E. Diacetylene linked anthracene oligomers synthesized by one-shot homocoupling of trimethylsilyl on Cu(111). *ACS Nano*, **12**, 8791−8797 (2018).

25. Sun, Q., Zhang, C., Li, Z., Kong, H., Tan, Q., Hu, A. & Xu, W. On-surface formation of one-dimensional polyphenylene through Bergman cyclization. *J. Am. Chem. Soc.*, **135**, 8448−8451 (2013).

26. Schuler, B., Fatayer, S., Mohn, F., Moll, N., Pavliček, N., Meyer, G., Peña, D. P. & Gross, L. Reversible Bergman cyclization by atomic manipulation. *Nat. Chem.*, **8**, 220−224 (2016).

27. Kanuru, V. K., Kyriakou, G., Beaumont, S. K., Papageorgiou, A. C., Watson, D. J. & Lambert, R. M. Sonogashira coupling on an extended gold surface in vacuo: reaction of phenylacetylene with iodobenzene on Au(111). *J. Am. Chem. Soc.*, **132**, 8081–8086 (2010).

28. Wang, T., Huang, J., Lv, H., Fan, Q., Feng, L., Tao, Z., Ju, H., Wu, X., Tait, S. L. & Zhu, J. Kinetic strategies for the formation of graphyne nanowires via Sonogashira coupling on Ag(111). *J. Am. Chem. Soc.*, **140**, 13421−13428 (2018).

29. Sun, K., Sagisaka, K., Peng, L., Watanabe, H., Xu, F., Pawlak, R., Meyer, E., Okuda, Y., Orita, A. & Kawai, S. Head-to-tail oligomerization by silylene-tethered Sonogashira coupling on Ag(111). *Angew. Chem. Int. Ed.*, **60**, 19598−19603 (2021).

30. Zhong, D., Franke, J. H., Podiyanachari, S. K., Blomker, T., Zhang, H., Kehr, G., Erker, G., Fuchs, H. & Chi, L. Linear alkane polymerization on a gold surface. *Science*, **334**, 213−216 (2011).

31. Sun, K., Chen, A., Liu, M., Zhang, H., Duan, R., Ji, P., Li, L., Li, Q., Li, C., Zhong, D., Müllen, K. & Chi, L. Surface-assisted alkane polymerization: investigation on structure−reactivity relationship. *J. Am. Chem. Soc.*, **140**, 4820−4825 (2018).

32. Sun, K., Silveira, O. J., Ma, Y., Hasegawa, Y., Matsumoto, M., Kera, S., Krejčí, O., Foster, A. & S., Kawai, S. On-surface synthesis of disilabenzene-bridged covalent organic frameworks. *Nat. Chem.*, **15**, 136–142 (2023).

33. Deniz, O., Sánchez-Sánchez, C., Dumslaff, T., Feng, X., Narita, A., Müllen, K., Kharche, N., Meunier, V., Fasel, R. & Ruffieux, P. Revealing the electronic structure of silicon intercalated armchair graphene nanoribbons by scanning tunneling spectroscopy. *Nano Lett.*, **17**, 2197 −2203 (2017).

34. Sun, K. & Kawai, S. Strength of electronic decoupling of fullerene on an AuSi$_X$ layer

formed on Au(111). *Phys. Chem. Chem. Phys.*, **23**, 5455−5459 (2021).

35. Sun, K., Nishiuchi, T., Sahara, K., Kubo, T., Foster, A. S. & Kawai, S. Low-temperature removal of dissociated bromine by silicon atoms for an on-surface Ullmann reaction. *J. Phys. Chem. C*, **124**, 19675−19680 (2020).

36. Gross, L., Mohn, F., Moll, N., Liljeroth, P. & Meyer, G. The chemical structure of a molecule resolved by atomic force microscopy. *Science*, **325**, 1110−1114 (2009).

37. Temirov, R., Soubatch, S., Neucheva, O., Lassise, A. C. & Tautz, F. S. A novel method achieving ultra-high geometrical resolution in scanning tunnelling microscopy. *New J. Phys.*, **10**, 053012 (2008).

38. Pyykkö, P. Additive covalent radii for single-, double-, and triple-bonded molecules and tetrahedrally bonded crystals: a summary. *J. Phys. Chem. A*, **119**, 2326−2337 (2015).

39. Sun, K., Li, X., Chen, L., Zhang, H. & Chi, L. Substrate-controlled synthesis of 5-armchair graphene nanoribbons. Substrate-controlled synthesis of 5-armchair graphene nanoribbons. *J. Phys. Chem. C*, **124**, 11422−11427 (2020).

40. Zhang, H., Lin, H., Sun, K., Chen, L., Zagranyarski, Y., Aghdassi, N., Duhm, S., Li, Q., Zhong, D., Li, Y., Müllen, K., Fuchs, H. & Chi, L. On-surface synthesis of rylene-type graphene nanoribbons. *J. Am. Chem. Soc.* 2015, **137**, 4022−4025 (2015).

41. Sheehan, W. F. & Schomaker, V. The Si-C bond distance in $Si(CH_3)_4$. *J. Am. Chem. Soc.*, **74**, 3956−3957 (1952).

42. Goel, S., Luo, X. & Reuben, R. L. Shear instability of nanocrystalline silicon carbide during nanometric cutting. *Appl. Phys. Lett.*, **100**, 231902 (2012).

43. Neaton, J. B., Hybertsen, M. S. & Louie, S. G. Renormalization of molecular electronic levels at metal-molecule interfaces. *Phys. Rev. Lett.*, **97**, 216405 (2006).

44. Thygesen, K. S. & Rubio, A. Renormalization of molecular quasiparticle levels at metal-molecule interfaces: trends across binding regimes, *Phys. Rev. Lett.*, **102**, 046802 (2009).

45. Rajca, A., Safronov, S., Rajca, S., Ross, R. C. & Stezowski, J. J. *J. Am. Chem. Soc.*, **118**, 7272-7279 (1996).

46. Krejčí, O., Hapala, P., Ondráček, M. & Jelínek, P. Principles and simulations of high-resolution STM imaging with a flexible tip apex. *Phys. Rev. B*, **95**, 045407 (2017).

47. Cai, S., Kurki, L., Xu, C., Foster, A. S. & Liljeroth, P. Water dimer-driven DNA base superstructure with mismatched hydrogen bonding. *J. Am. Chem. Soc.*, **144**, 20227–20231 (2022).

48. Kokalj, A. XCrySDen–a new program for displaying crystalline structures and electron densities. *J. Mol. Graphics Modelling*, **17**, 176–179 (1999).

49. Humphrey, W., Dalke, A. & Schulten, K. VMD: visual molecular dynamics. *J. Molec. Graphics*, **14**, 33–38 (1996).

50. Momma, K. & Izumi, F. VESTA: a three-dimensional visualization system for electronic and structural analysis. *J. Appl. Cryst.*, **41**, 653–658 (2008).

# Supporting Information for
# On-Surface Synthesis of Silole and Disilacyclooctaene Derivatives

Kewei Sun[1]*, Lauri Kurki[3], Orlando J. Silveira[3], Tomohiko Nishiuchi[4], Takashi Kubo[4], Ondřej Krejčí[3], Adam S. Foster[3,5]*, Shigeki Kawai[2,6]*

[1]*International Center for Young Scientists, National Institute for Materials Science, 1-2-1 Sengen, Tsukuba, Ibaraki 305-0047, Japan.*

[2]*Center for Basic Research on Materials, National Institute for Materials Science, 1-2-1 Sengen, Tsukuba, Ibaraki 305-0047, Japan.*

[3]*Department of Applied Physics, Aalto University, Espoo, Finland*

[4]*Department of Chemistry, Graduate School of Science, Osaka University, Toyonaka 560-0043, Japan.*

[5] *Nano Life Science Institute (WPI-NanoLSI), Kanazawa University, Kakuma-machi, Japan*

[6]*Graduate School of Pure and Applied Sciences, University of Tsukuba, Tsukuba 305-8571, Japan.*

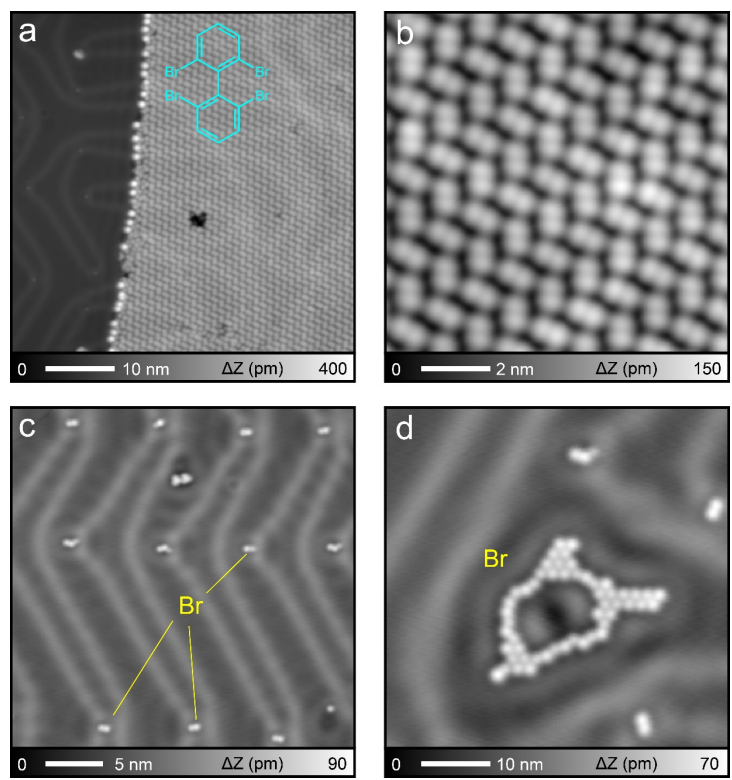

**Figure S1. Molecules 1 on Au(111).** (**a**) STM topography of as-deposited **1** on Au (111) surface at room temperature. (**b**) The close-up of self-assembled **1**. (**c,d**) STM topographies of Au (111) surface after annealing at 420 K for 5 min. Molecules **1** have disappeared. Some Br atoms are visible, indicating that debromination of some **1** took place. Measurement parameters: Sample bias voltage $V$ = 200 mV and tunneling current $I$ = 5 pA in (a). $V$ = 50 mV and $I$ = 10 pA in (b). $V$ = 100 mV and $I$ = 10 pA in (c). $V$ = 200 mV and $I$ = 10 pA in (d).

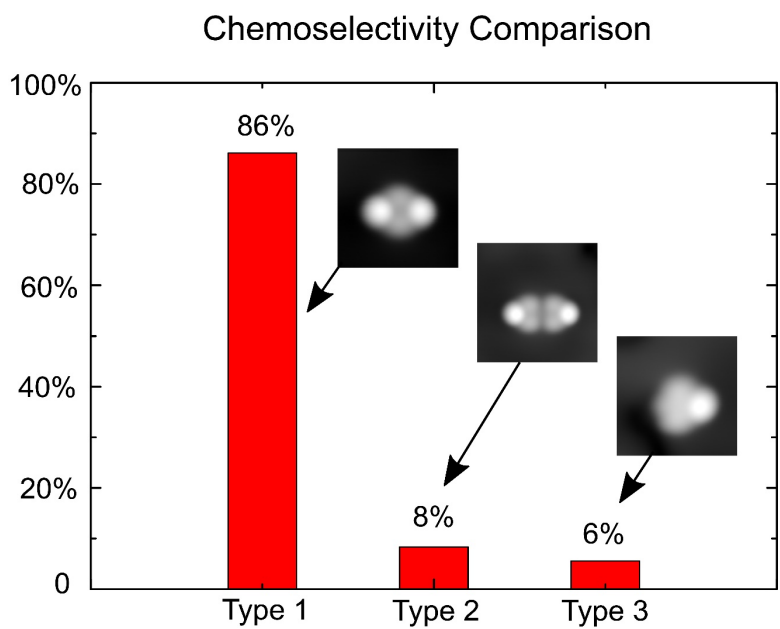

**Figure S2. Statistical analysis of chemoselectivity comparison for three types of products.**

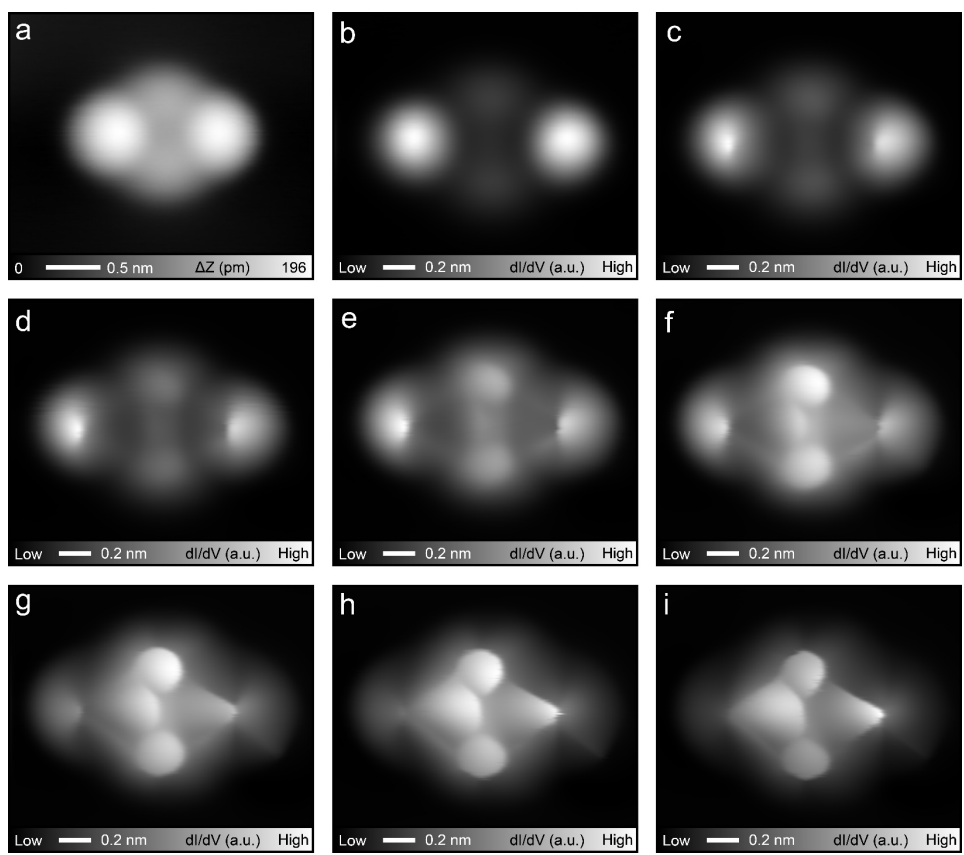

**Figure S3. High-resolution images of Type 1.** (**a**) STM topography of one Type 1, which is identical with that in Figure 1b of the main text. (**b-i**) A series of bond-resolved constant height d$I$/d$V$ maps recorded by a CO-tip with decreasing tip-surface distances. Measurement parameters: $V$ = 200 mV and $I$ = 5 pA in (a).

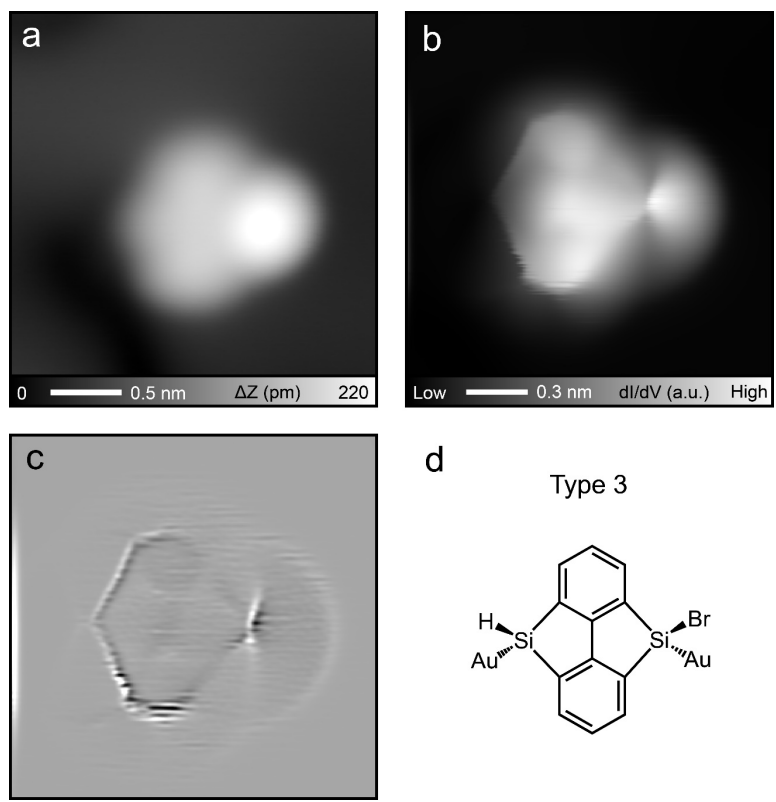

**Figure S4. Type 3 on Au(111).** (**a**) STM topography of one Type 3. (**b**) Constant height d$I$/d$V$ map of the Type 3 in (a), and (**c**) the corresponding Laplace filtered image. (**d**) The possible chemical structure of Type 3. Measurement parameters: $V$ = 200 mV and $I$ = 5 pA in (a).

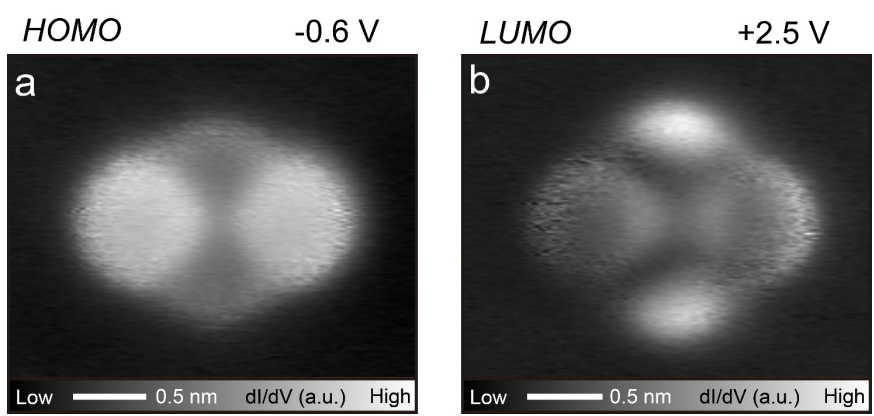

**Figure S5. STS measurement on Type 1.** Constant current d$I$/d$V$ maps measured at bias voltages of (**a**) -0.6 V and (**b**) 2.5 V. Measurement parameters: $V$ = -0.6 V, $I$ = 200 pA, $V_{ac}$ = 10 mV in (a). $V$ = 2.5 V, $I$ = 600 pA, $V_{ac}$ = 10 mV in (b).

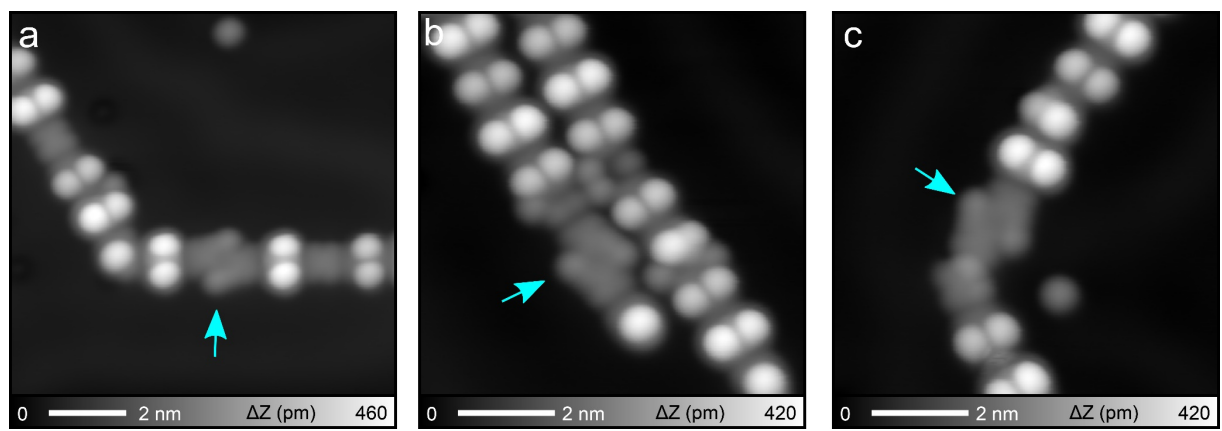

**Figure S6. More GNRs on Au(111).** (**a-c**) A series of STM topographies of eight-membered sila-cyclic rings bridged GNRs with similar segment (indicated by arrows). Measurement parameters: $V$ = 200 mV and $I$ = 5 pA.

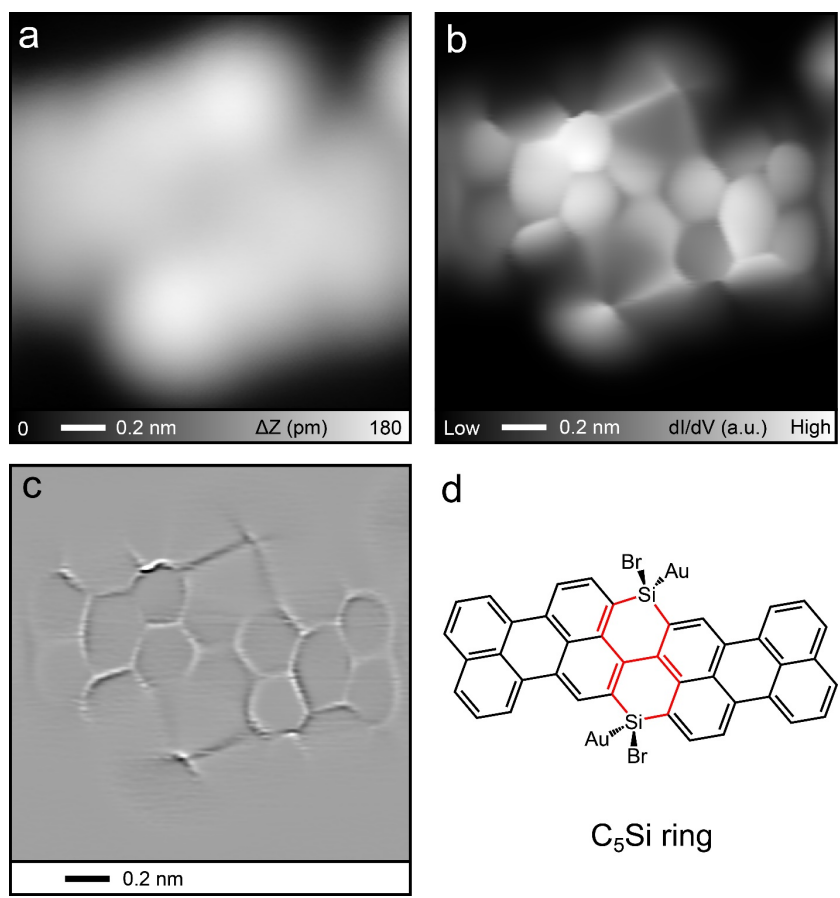

**Figure S7. C$_5$Si rings on Au(111).** (**a**) Close-up of the segment. It is identical with the inset of Figure 5a. (**b**) Constant height d$I$/d$V$ map, which is the same as Figure 5b of the main text and (**c**) the Laplace filtered image, and (**d**) the corresponding chemical structure. Measurement parameters: $V$ = 200 mV and $I$ = 5 pA in (**a**).

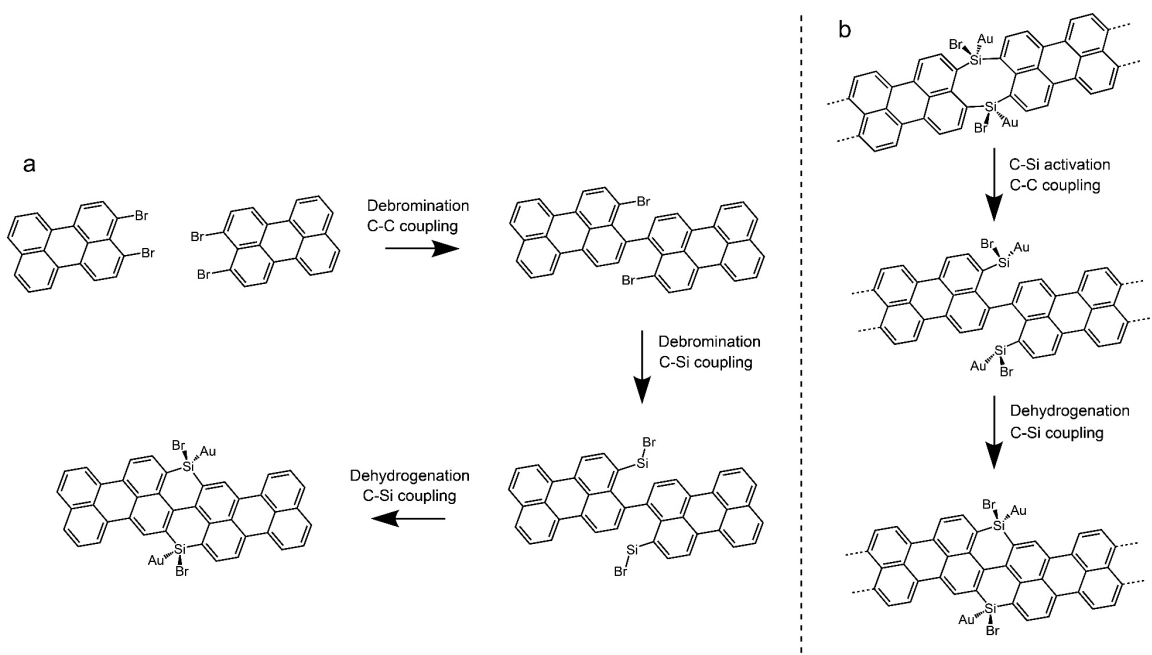

**Figure S8.** Proposed reaction pathways for the formation of two $C_5Si$ cyclic rings incorporated GNR segment.

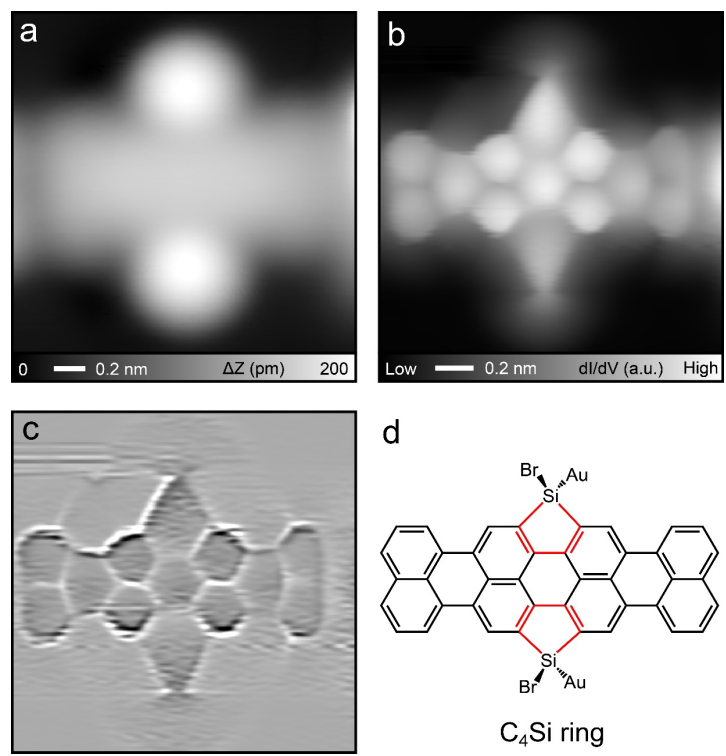

**Figure S9. $C_4Si$ rings on Au(111).** (**a**) STM topography of GNR segment with two dots at edges, which is the same as the inset of Figure 5d of the main text. (**b**) Constant height d$I$/d$V$ map, which is the same as the inset of Figure 5e, (**c**) the corresponding Laplace filtered image, and (**d**) the chemical structure. Measurement parameters: $V = 200$ mV and $I = 5$ pA in (a).

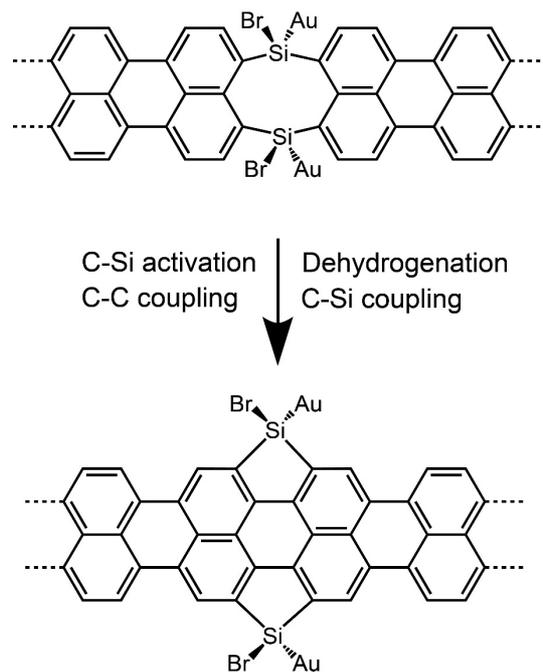

**Figure S10.** Proposed reaction pathway for the formation of two C$_4$Si cyclic rings incorporated GNR segment.

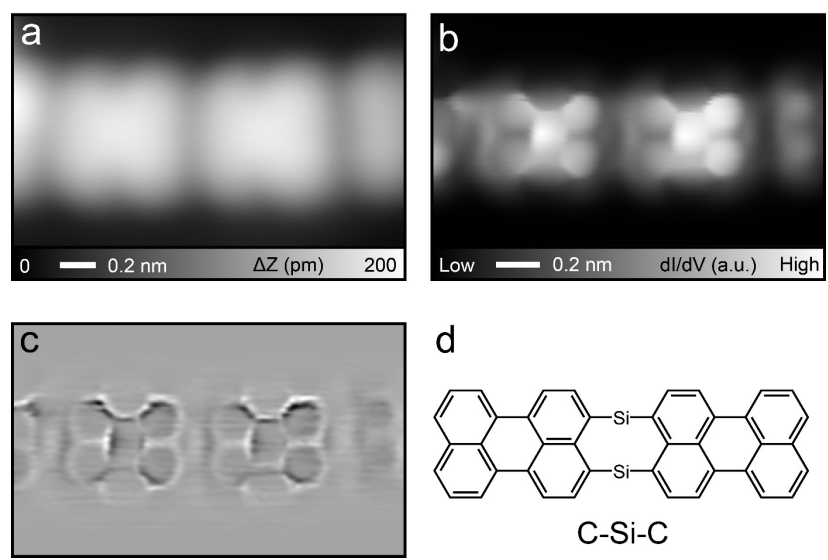

**Figure S11. GNR segment on Au(111).** (**a**) STM topography of GNR segment with dark line, which is the same as the inset of Figure 5g of the main text. (**b**) Constant height d$I$/d$V$ map, which is the same as Figure 5h of the main text. (**c**) The corresponding Laplace filtered image. (**d**) The proposed chemical structure after debromination of C$_6$Si$_2$Br$_2$ moieties between two perylene. Measurement parameters: $V = 200$ mV and $I = 5$ pA in (a).

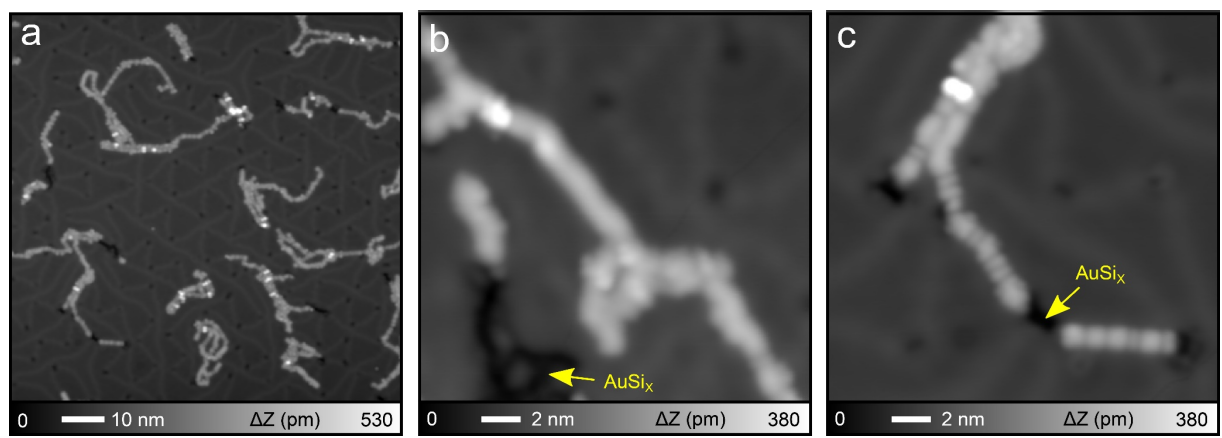

**Figure S12. A series of STM topographies of the sample after annealing at 620 K for 5 min.** (**a**) Large-scale STM image. (**b**) STM image of one long GNR and some segments. (**c**) STM image of GNR with many dark lines. Several dark AuSi$_X$ areas (indicated by arrows) can be identified, in which Si are probably desorbed from sila-cyclic rings after annealing a high temperature. Measurement parameters: $V$ = 200 mV and $I$ = 5 pA in (a). $V$ =100 mV and $I$ = 20 pA in (b). $V$ = 300 mV and $I$ = 20 pA in (c).

**DFT calculations for isolated type 1 and type 2**

We performed further DFT calculations of the isolated type **1** molecule at the B3LYP[1,2] level as implemented as in ORCA[3] code, where we employed the def2-SPV[4] basis set. At first, we considered planar geometry, and buckled configuration as can be seen in Fig S13. Additionally, we considered the possibility of closed shell singlet and triplet solutions. Our calculations show that the buckling lowers down significantly the total energy of the molecule on both singlet (102 meV) and triplet (647 meV) cases. Furthermore, unpaired electrons arising from the Si atoms result in a triplet configuration 77 meV lower in energy than the closed shell solution for the planar case, while in the buckled configuration this difference in total energy drastically lowers to –1.3 eV. In the on-surface case, on the other hand, gold atoms are available in the substrate to passivate the unpaired electrons, bringing the molecule to a closed shell solution as a Si-Au bond is created. Fig S13 reveals that the energy is lowered by the single occupation of the closed shell HOMO and LUMO, which also significantly increases the band gap of the molecule.

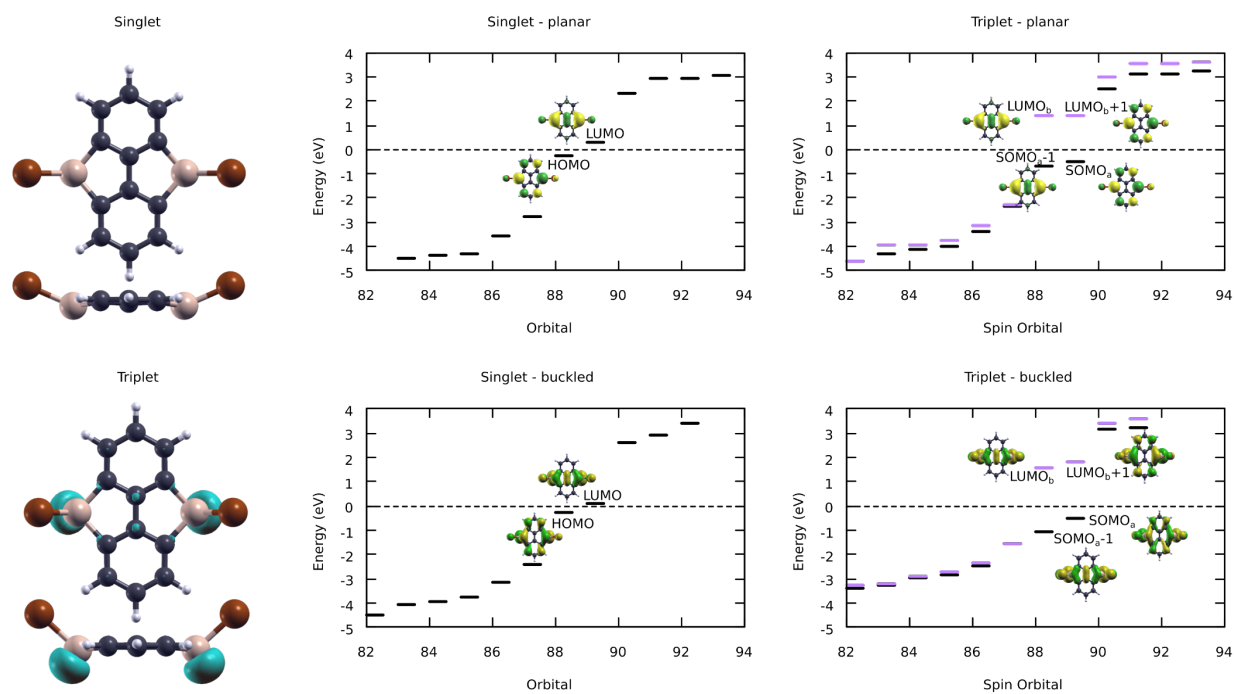

**Figure S13.** The geometries on the left correspond to the top and side view of the type 1 molecule, obtained considering a singlet solution (top) and a triplet solution (bottom). The geometry of the triplet has also the isosurface of the spin density drawn at 0.001 e bohr$^{-3}$. The energy diagrams on the right correspond to the orbital energies of the planar and buckled geometries (top and bottom), both considering the singlet and triplet solutions (middle and right). The insets are selected molecular or spin orbitals, drawn at 0.03 e bohr$^{-3}$.

**References**

[1] Becke, A.D. *J. Chem. Phys.*, **98**, 5648–5652 (1993).

[2] Stephens, P.J.; Devlin, F.J.; Chabalowski, C.F. & Frisch, M.J. *J. Phys. Chem.*, **98**, 11623-11627 (1994).

[3] Neese, F. *WIREs Comput. Mol. Sci.*, **2**, 73–78 (2012).

[4] Zheng, J.; Xu, X. & Truhlar, D.G. *Theor. Chem. Acc.*, **128**, 295–305 (2011).